\documentclass[sigplan,10pt,nonacm]{acmart}
\usepackage{xcolor}
\usepackage{xspace}
\usepackage{comment}
\usepackage{booktabs}
\usepackage{listings}
\usepackage{makecell}
\usepackage{etoolbox}
\usepackage{url}
\usepackage{hyperref}
\usepackage{amsmath,amsfonts}
\usepackage{graphicx}
\usepackage{textcomp}
\usepackage{tikz}
\usepackage{caption}
\usepackage{subcaption}
\usepackage{balance}
\usepackage[a-2b]{pdfx}

\usepackage{enumitem}
\setlist{nolistsep}

\newtoggle{showmarks}
\toggletrue{showmarks}

\iftoggle{showmarks}{
  \newcommand\myzhao[1]{\textcolor{violet}{[~MARK:~#1~]}}
  \newcommand\todo[1]{\textcolor{red}{[~TODO:~#1~]}}
  \newcommand\remark[1]{\textcolor{orange}{[~REMARK:~#1~]}}
  \newcommand\christos[1]{\textcolor{blue}{[~Christos:~#1~]}}
}{
  \newcommand\myzhao[1]{\unskip}
  \newcommand\todo[1]{\unskip}
  \newcommand\remark[1]{\unskip}
  \newcommand\christos[1]{\unskip}
}

\newcommand{\systemname}{\textsc{DataHelix}\xspace}

\definecolor{codegreen}{rgb}{0,0.6,0}
\definecolor{codegray}{rgb}{0.5,0.5,0.5}
\definecolor{codepurple}{rgb}{0.58,0,0.82}
\definecolor{backcolour}{rgb}{0.95,0.95,0.92}

\lstdefinestyle{python}{
    backgroundcolor=\color{backcolour},   
    commentstyle=\color{codegreen},
    keywordstyle=\color{magenta},
    numberstyle=\tiny\color{codegray},
    stringstyle=\color{codepurple},
    basicstyle=\ttfamily\footnotesize,
    breakatwhitespace=false,         
    breaklines=true,                 
    captionpos=b,                    
    keepspaces=true,                 
    numbers=left,                    
    numbersep=5pt,                  
    showspaces=false,                
    showstringspaces=false,
    showtabs=false,                  
    tabsize=2
}

\usepackage{algorithm}
\usepackage{algpseudocode}
\usepackage{amsmath}
\usepackage{booktabs}
\usepackage{makecell}
\usepackage{multirow}
\usepackage{siunitx}
\usepackage{array}
\usepackage{xcolor}
\usepackage{listings}
\usepackage[scaled=0.85]{beramono}

\definecolor{codegray}{gray}{0.95}
\definecolor{commentgray}{gray}{0.45}

\lstdefinestyle{pythoncompact}{
    language=Python,
    basicstyle=\ttfamily\scriptsize,
    numbers=left,
    numberstyle=\tiny,
    numbersep=4pt,
    frame=single,
    breaklines=true,
    columns=fullflexible,
    keepspaces=true,
    showstringspaces=false,
    tabsize=2,
    captionpos=b
}

\newcolumntype{L}[1]{>{\raggedright\arraybackslash}m{#1}}

\setcopyright{none}
\makeatletter
\newcommand{\preprintauthors}{}
\newcommand{\preprintauthor}[2]{%
  \author{#1}%
  \ifdefempty{\preprintauthors}{}{\gappto\preprintauthors{\quad}}%
  \gappto\preprintauthors{#1\textsuperscript{#2}}}
\renewcommand{\@mkauthors}{%
  \begingroup
  \hsize=\textwidth
  \global\setbox\mktitle@bx=\vbox{%
    \noindent\box\mktitle@bx\par\medskip
    \centering
    {\large\normalfont\preprintauthors\par}
    \medskip
    {\normalsize\normalfont
      \textsuperscript{1}Shanghai Jiao Tong University\quad
      \textsuperscript{2}Xiamen University\par}
    \bigskip}%
  \endgroup}
\makeatother

\begin{document}
\title{\systemname: 
A CPU-GPU Co-Scheduling Data Plane for Multimodal Model Training
}

\preprintauthor{Yuanhe Li}{1}
\affiliation{
  \institution{Shanghai Jiao Tong University}
}

\preprintauthor{Jiazhen Wu}{2}
\affiliation{
  \institution{Xiamen University}
}

\preprintauthor{Dian Ding}{1}
\affiliation{
  \institution{Shanghai Jiao Tong University}
}

\preprintauthor{Guangtao Xue}{1}
\affiliation{
  \institution{Shanghai Jiao Tong University}
}

\preprintauthor{Yiming Zhang}{1}
\affiliation{
  \institution{Shanghai Jiao Tong University}
}

\begin{abstract}

Multimodal model training breaks the loose coupling between data loading and model execution assumed by conventional training pipelines. Large per-sample variation in decoding cost, token/latent count, and tensor size causes both CPU preprocessing time and GPU compute time to vary substantially across data-parallel ranks (training processes). In synchronous training, these two sources of heterogeneity interact tightly: GPU execution determines how long the input pipeline has to prepare the next batch, while delayed preprocessing in any rank in turn stalls all GPUs at synchronization. As a result, optimizing CPU throughput or GPU load balance in isolation cannot eliminate pipeline bubbles.

We present \systemname, a metadata-first CPU–GPU co-scheduling data plane for multimodal training. DataHelix jointly reasons about preprocessing demand and GPU execution progress, treating each GPU compute window as the readiness budget for the next batch and coordinating preprocessing across ranks to minimize deadline misses rather than CPU makespan. To avoid expensive data movement, placement is decided while samples remain lightweight descriptors, before decoding and collation amplify them into large tensors. Descriptor-level token-/latent-aware batching performs coarse pre-balancing, while cost-gated collective rebalancing corrects residual skew only when its expected benefit exceeds transfer cost.
Across 11 video and multimodal workloads on 8 Ascend 910B accelerators, \systemname reduces mean deadline-miss rate by 67.8\% and mean p95 input-wait latency by 87.0\%, while achieving 2.72$\times$ geometric-mean throughput over rank-local DataLoaders and 1.82$\times$ over a Ray actor pool under the same CPU budget.

\end{abstract}

\pagestyle{plain}
\maketitle



\section{Introduction}\label{sec:introduction}
\begin{figure}[t]
  \centering
  \includegraphics[width=\linewidth]{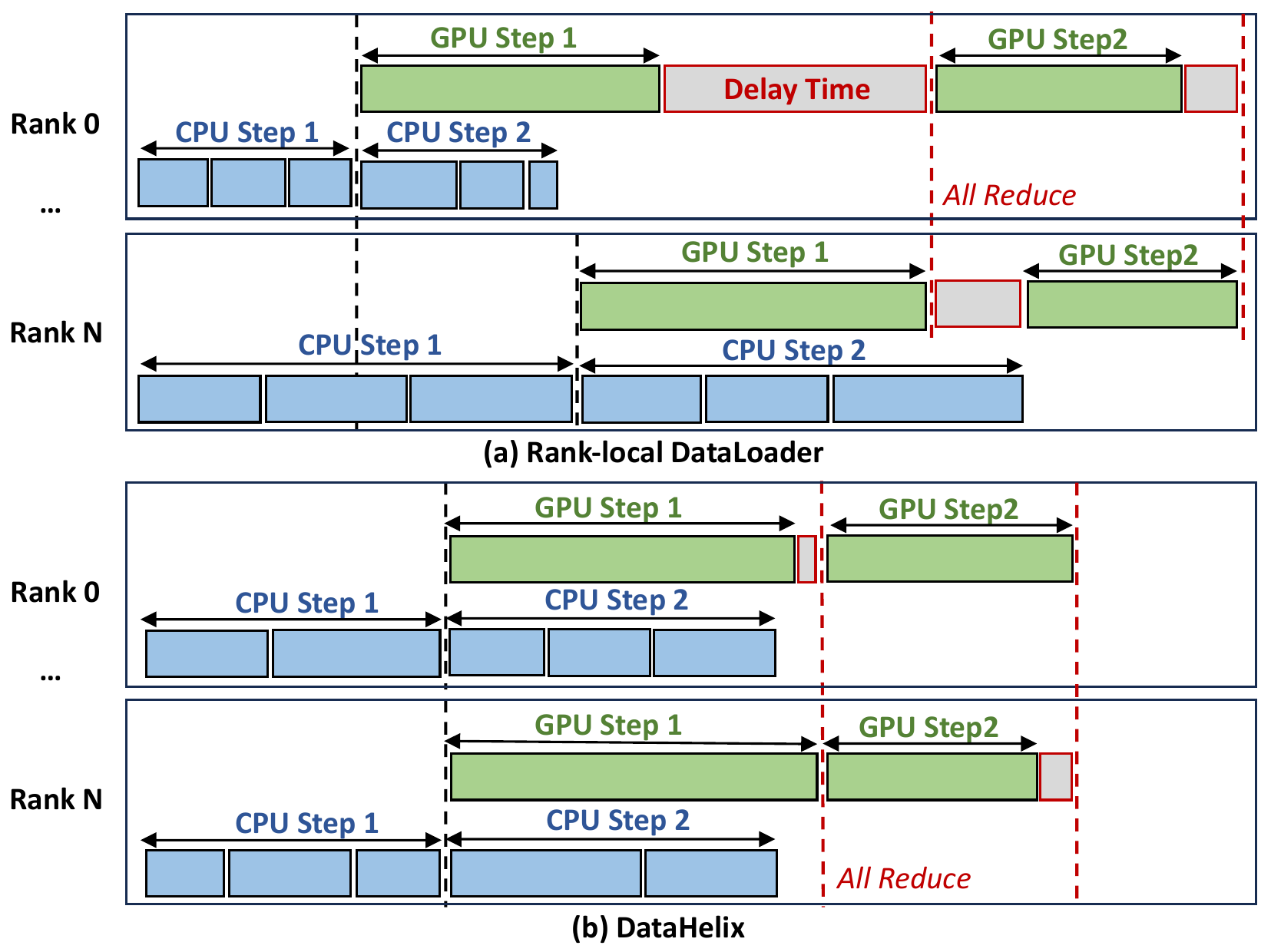}
  \caption{CPU--GPU coupling: preprocessing must meet per-rank readiness deadlines set by GPU execution.}
  \label{fig:simclr}
\end{figure}

Conventional training input pipelines assume sample count proxies computational cost, so equal sample distribution suffices for load balance. That holds for homogeneous text or image workloads but breaks down entirely in multimodal large language model (MLLM) training such as long-video understanding, VLM pretraining, and diffusion transformer (DiT) fine-tuning. Within one batch, preprocessing costs vary by orders of magnitude: CPU decoding latency scales superlinearly with video duration and resolution, spatial transformation cost grows with pixel count, and the resulting sequence lengths—visual tokens in autoregressive VLMs or spatial latents in DiT architectures—diverge by 10–100×. Equal sample count no longer implies equal preprocessing time, much less equal GPU compute time. The imbalance has a global consequence: in synchronous data-parallel training, the slowest rank's preprocessing delay determines when all ranks advance, leaving faster ranks idle at the gradient synchronization barrier while a single straggler drags down cluster-wide GPU utilization—not merely a local CPU preprocessing issue but an architectural problem cascading into global GPU efficiency.

\begin{figure}[t]
  \centering
  \begin{subfigure}{\linewidth}
    \centering
    \includegraphics[width=\linewidth]{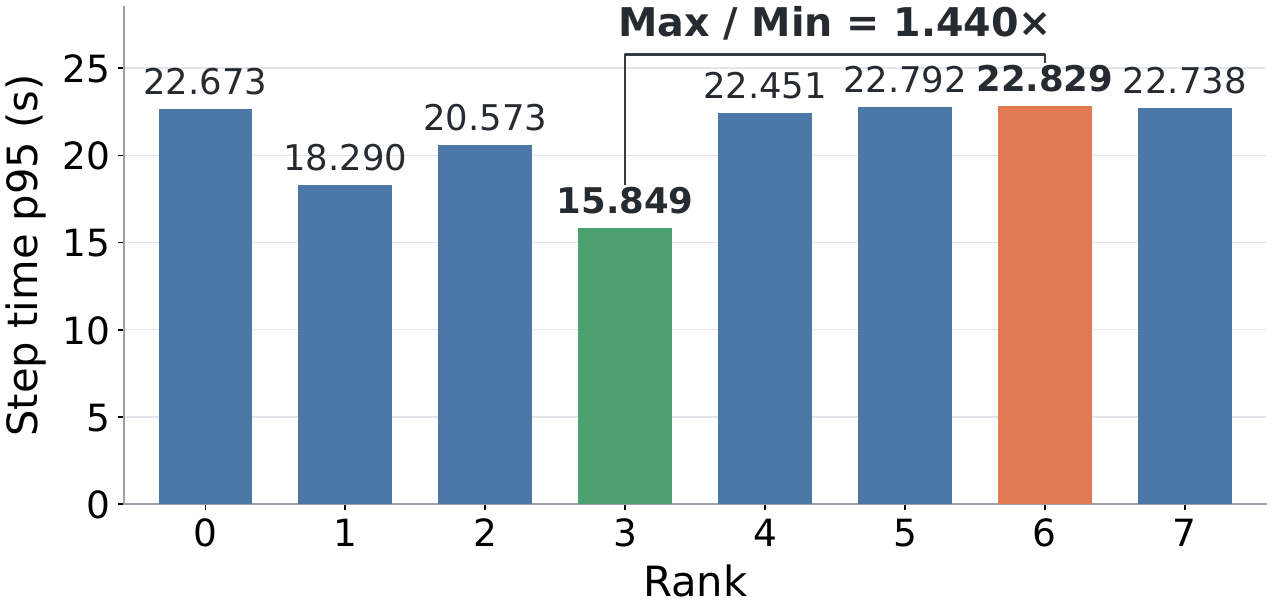}
    \caption{Per-rank step-time p95.}
    \label{fig:native_rank_p95}
  \end{subfigure}
  \par\smallskip
  \begin{subfigure}{\linewidth}
    \centering
    \includegraphics[width=\linewidth]{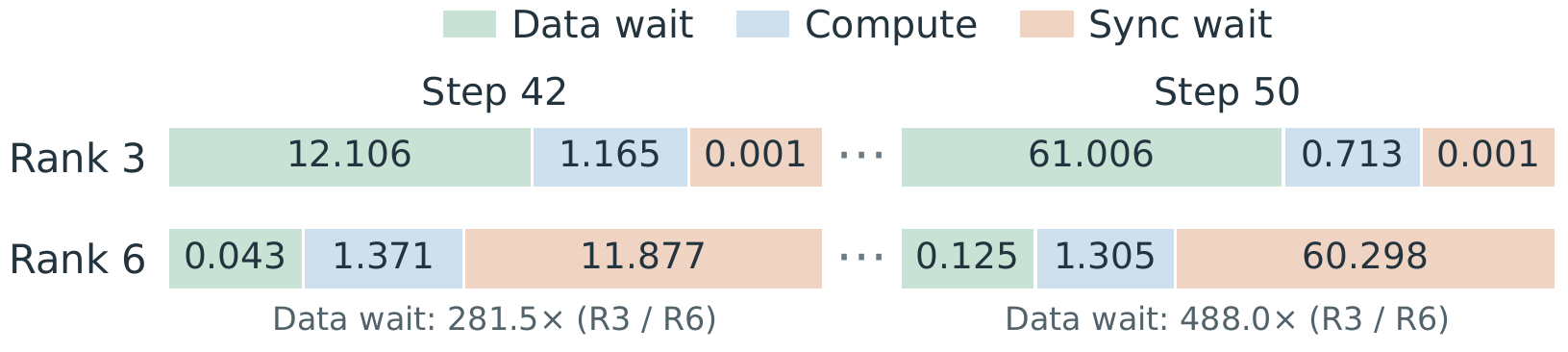}
    \caption{Selected steps (s; widths not to scale).}
    \label{fig:native_step_queue}
  \end{subfigure}
  \caption{Qwen-VL rank skew and long waits in native synchronous training (batch size 2, eight ranks).}
  \label{fig:native_rank_motivation}
  \Description{The upper panel shows Qwen-VL step-time p95 across eight ranks, with a maximum-to-minimum ratio of 1.44. The lower panel shows separately supplied Qwen-VL steps 42 and 50 with long data waits on Rank 3 and long synchronization waits on Rank 6. Segment widths are schematic; normal-weight labels report seconds.}
\end{figure}

Figure~\ref{fig:native_rank_motivation} shows cross-rank latency variation and long-tail waits in native synchronous Qwen-VL training with batch size 2 and eight ranks. Per-rank step-time p95 ranges from 15.849\,s to 22.829\,s, a $1.44\times$ gap. Separately supplied step measurements provide a complementary view: in Step~50, Rank~3 waits 61.006\,s for data, while Rank~6 spends 60.298\,s in synchronization. These measurements motivate coordinating batch readiness across ranks to limit input stalls and synchronization waiting, rather than optimizing aggregate preprocessing throughput alone.


Fundamentally, multimodal heterogeneity tightly couples CPU data production with GPU execution. While the GPU processes batch \(t\), the input pipeline has only that compute window to prepare batch \(t+1\); conversely, if any rank falls behind, the resulting data stall propagates to all ranks through synchronization. GPU progress therefore determines the preprocessing budget, while preprocessing readiness determines whether GPU execution can advance. Figure~\ref{fig:latency_breakdown} makes this coupling visible in multimodal VLM training: under rank-local DataLoaders, data wait and synchronization wait account for a substantial fraction of the summed rank-0 component summaries. These summaries diagnose waiting behavior rather than measured end-to-end step latency. The relevant objective is thus not aggregate preprocessing throughput or CPU makespan, but keeping per-rank data production aligned with the GPU execution timeline.

Prior work supplies important building blocks. DynaPipe~\cite{jiang2024dynapipe} adapts microbatches and pipeline execution to variable sequence lengths. tf.data service~\cite{audibert2022tfdata_service} disaggregates preprocessing, Pecan~\cite{graur2024pecan} dynamically places workers across local and remote CPUs, and MegaScale-Data~\cite{zhao2026megascale} coordinates multisource loading under heterogeneous preprocessing costs. Thus, remote execution and global coordination are established techniques; our focus is coupling input placement to each rank's batch-readiness target.
This coupling requires accounting for when and where data becomes expensive to move. FastFlow~\cite{um2023fastflow} already considers network bandwidth in offloading decisions. For heterogeneous multimodal batches, we jointly consider candidate-specific preprocessing and delivery costs against the consuming rank's GPU window, make initial assignments before tensor materialization, and gate later corrections by their transfer cost.

Coordinating these decisions calls for \textbf{a CPU–GPU Co-Scheduling Data Plane}: a globally coordinated substrate that uses the GPU compute window to set a per-step readiness target, decides placement and routing at the descriptor stage before any tensor is materialized, and jointly optimizes CPU allocation and inter-node transfer cost across all ranks. Realizing this vision surfaces four interrelated challenges.

\textbf{Challenge 1: Preprocessing cost cannot be known before execution.}
Token and latent counts, CPU decoding latency, and output tensor sizes are unknown until preprocessing completes, yet decisions must be made at the descriptor stage before any tensor is materialized. A scheduler relying on static estimates risks systematic mis-scheduling; waiting for exact measurements adds unacceptable latency.

\textbf{Challenge 2: Placement must track synchronized consumption.}
Work stealing~\cite{blumofe1999scheduling} and heterogeneous-cluster straggler mitigation~\cite{zaharia2008improving} improve task completion. For training inputs, placement must also consider when each rank consumes its batch: a lightly loaded remote worker may still deliver too late after decoding and tensor transfer. A readiness miss can then propagate through synchronization, motivating per-rank deadline scoring alongside load and locality.

\textbf{Challenge 3: Data amplification makes post-materialization migration prohibitively expensive.}
A sample occupies negligible storage as a descriptor but expands by orders of magnitude after decoding and collation: a short clip may take kilobytes at the index stage yet produce gigabytes of decoded RGB tensor. Deferring cross-node rebalancing until after materialization can cost more in transfer than re-decoding locally, and the benefit threshold is non-monotone in both sample size and network conditions.

\textbf{Challenge 4: Integrating with the GPU execution loop without introducing synchronization overhead.}
Deadline-aware scheduling requires real-time feedback from the GPU execution layer—the remaining compute window and per-rank progress—yet coupling the data pipeline to the training process risks extra synchronization barriers or compromised fault isolation.

We present \textbf{\systemname}, \textbf{a CPU--GPU Co-Scheduling Data Plane} for multimodal foundation model training that reframes the input system as a cluster-level data plane using GPU compute windows to set per-rank readiness targets.

At its core, \systemname decouples control from data movement: the Coordinator makes placement and routing decisions at the descriptor stage before any tensor is materialized, while DataServers execute preprocessing and deliver payloads. A deadline-aware scheduler routes requests to minimize per-rank deadline misses, token/latent-aware batching equalizes GPU compute load, and cost-gated rebalancing corrects residual skew using collective communication. An asynchronous pipeline overlaps preprocessing with GPU execution without extra barriers. Together these mechanisms preserve standard DataLoader compatibility while addressing the architectural gap between rank-local loading and deadline-aware scheduling.

\systemname has been seamlessly integrated into Pytorch and MindSpore \cite{mindspore2020}. 
We make the following contributions.
\begin{itemize}[leftmargin=*]

    \item We identify CPU–GPU coupling as a key bottleneck in synchronous multimodal training and show that the relevant objective is per-rank data readiness along the GPU execution timeline, rather than aggregate CPU throughput. We present \systemname, a cluster-level CPU–GPU co-scheduling data plane that minimizes deadline misses across ranks.

    \item We propose a metadata-first, deadline-aware placement architecture that combines descriptor-level cost estimates with runtime CPU, network, and GPU feedback to coordinate preprocessing before tensor materialization.
     \item  We design a two-tier load-balancing mechanism that combines token/latent-aware pre-balancing with cost-gated HCCL/NCCL-based residual correction, together with an asynchronous pipeline that overlaps preprocessing with GPU execution without extra barriers.
    

    \item  We evaluate \systemname on an 8-card Ascend 910B server across 11 video and multimodal workloads, \systemname reduces mean deadline-miss rate by 67.8\% and mean p95 input-wait latency by 87.0\%, while improving geometric-mean throughput by 2.72× over rank-local DataLoaders and 1.82× over a Ray actor pool under the same CPU budget. Beyond controlled evaluation, DataHelix has been upstreamed into the data engine of a widely used open-source AI framework with millions of downloads and tens of thousands of community contributors, and reported production deployments across thousands of enterprise users provide further evidence of its practicality at scale.

\end{itemize}

\section{Background}\label{sec:background}

This section characterizes the multimodal training data path, analyzes limitations of existing input systems, and summarizes the problems \systemname addresses. Unlike homogeneous image workloads, multimodal training exhibits pronounced per-sample heterogeneity in preprocessing cost, token count, and tensor volume, making the input pipeline a first-class determinant of GPU utilization and training throughput; large-scale production studies similarly show that data storage and ingestion can dominate training efficiency~\cite{zhao2022dsi}.

\subsection{Data Path of Multimodal Training}

\begin{figure}[t]
  \centering
  \includegraphics[width=\linewidth]{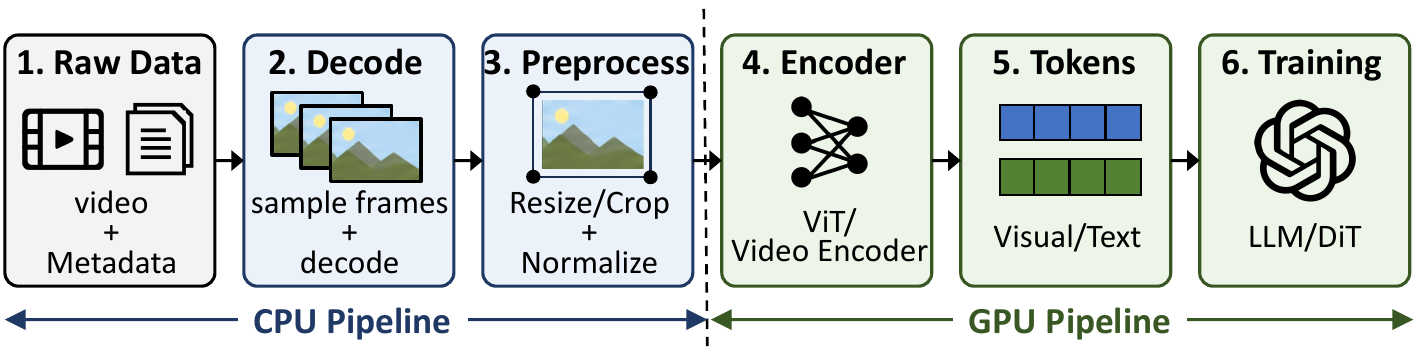}
  \vspace{-5mm}
  \caption{Data path of multimodal training.} 
  \label{fig:datapath}
\end{figure}

Framework loaders such as PyTorch DataLoader~\cite{paszke2019pytorch} and tf.data~\cite{murray2021tfdata} expose the transformations in Fig.~\ref{fig:datapath}. \texttt{Metadata probe} produces lightweight descriptors: duration, resolution, frame count, file path, and sampling plan. \texttt{Decode} and \texttt{resize/crop} consume CPU time according to codec, frame count, and resolution; \texttt{collate} and delivery depend on materialized tensor size. The visual encoder and downstream LLM or DiT dominate GPU compute. Preprocessing studies~\cite{isenko2022bottleneck} show why placement must account for these different resource demands.

\subsection{Limitations of Existing Input Systems}

\textbf{Observation 1: Sample count is no longer a valid proxy for workload.}
In conventional image classification, batch size approximates per-rank computation uniformly. In video, VLM, and DiT training, this assumption breaks down: a short clip may produce a handful of visual tokens, while a long or high-resolution sample may yield substantially more tokens, patches, or latent dimensions. Equal sample counts, therefore, translate to unequal GPU compute times, producing stragglers under synchronous training. CPU preprocessing cost exhibits the same heterogeneity—decode and transform latency varies with codec, frame count, resolution, and sampling strategy, and is no longer linear in sample count. 


\textbf{Observation 2: Batch readiness depends on the GPU compute window.}
Plumber~\cite{kuchnik2021plumber} and cedar~\cite{zhao2024cedar} optimize pipeline throughput; Cachew~\cite{graur2022cachew} uses worker and training-client feedback to avoid data stalls. Our scheduling target is \emph{per-rank batch readiness}: preparation and delivery of batch $t+1$ should fit within batch $t$'s GPU window to avoid a pipeline bubble:
$
t_{\mathrm{prep}}(\text{batch}_{t+1}) + t_{\mathrm{delivery}}(\text{batch}_{t+1}) \;\leq\; t_{\mathrm{GPU}}(\text{batch}_t)
$
Simple worker load balancing is insufficient: a DataServer with a short queue is not necessarily optimal if its I/O cost, output tensor size, or network path cannot satisfy the deadline. 


\textbf{Observation 3: Post-materialization rebalancing incurs prohibitive communication cost.}
Video samples undergo severe data amplification along the preprocessing pipeline: a descriptor occupies a few hundred bytes, a raw compressed video occupies megabytes, and a decoded, normalized tensor may reach tens to hundreds of megabytes. This asymmetry has a direct implication for scheduling: placement and routing decisions made at the descriptor stage incur negligible communication overhead, whereas load balancing deferred until after materialization may generate inter-node transfer costs that negate any straggler reduction benefit. This asymmetry demands that placement decisions be finalized at the descriptor stage and that any post-materialization correction be gated by a cost model.

\subsection{Problem Statement and Design Goals}

Based on the preceding analysis, we formalize the core problem: given a synchronous multi-GPU training job over heterogeneous multimodal samples, determine batch composition and DataServer placement at the descriptor stage—before any tensor is materialized—such that (1) each GPU rank receives an approximately equal token/latent compute load; (2) the full preprocessing pipeline for batch $t+1$ completes within the GPU compute window of batch $t$; (3) post-materialization rebalancing executes only when its projected straggler reduction exceeds inter-node transfer cost; and (4) the system preserves standard DataLoader compatibility while reducing communication overhead and improving end-to-end throughput.

These constraints map directly to three design goals: G1 addresses constraint (4) by decoupling control from data movement; G2 addresses constraint (2) through deadline-aware placement; and G3 jointly addresses constraints (1) and (3) via two-tier load balancing.
\textbf{G1 (High-performance data plane):} Decouple control from data movement—the Coordinator handles only lightweight metadata while DataServers execute preprocessing and deliver payloads via shared memory, streaming, or peer transfer—eliminating serialization overhead on the critical path while preserving standard DataLoader compatibility.
\textbf{G2 (Deadline-aware CPU scheduling):} Select DataServer placement by estimating end-to-end completion time per candidate, accounting for queue wait, I/O, CPU, output size, and network conditions, with all routing finalized at the descriptor stage before materialization.
\textbf{G3 (Two-tier GPU-side load balancing):} Combine token/latent-aware batch construction—a zero-migration coarse pre-balance at the descriptor stage—with cost-gated post-materialization rebalancing (DDLB) that corrects residual rank skew using true sample statistics only when projected straggler reduction exceeds inter-node transfer cost.

\section{\systemname Design}\label{sec:design}

\begin{figure}[t]
  \centering
  \includegraphics[width=\linewidth]{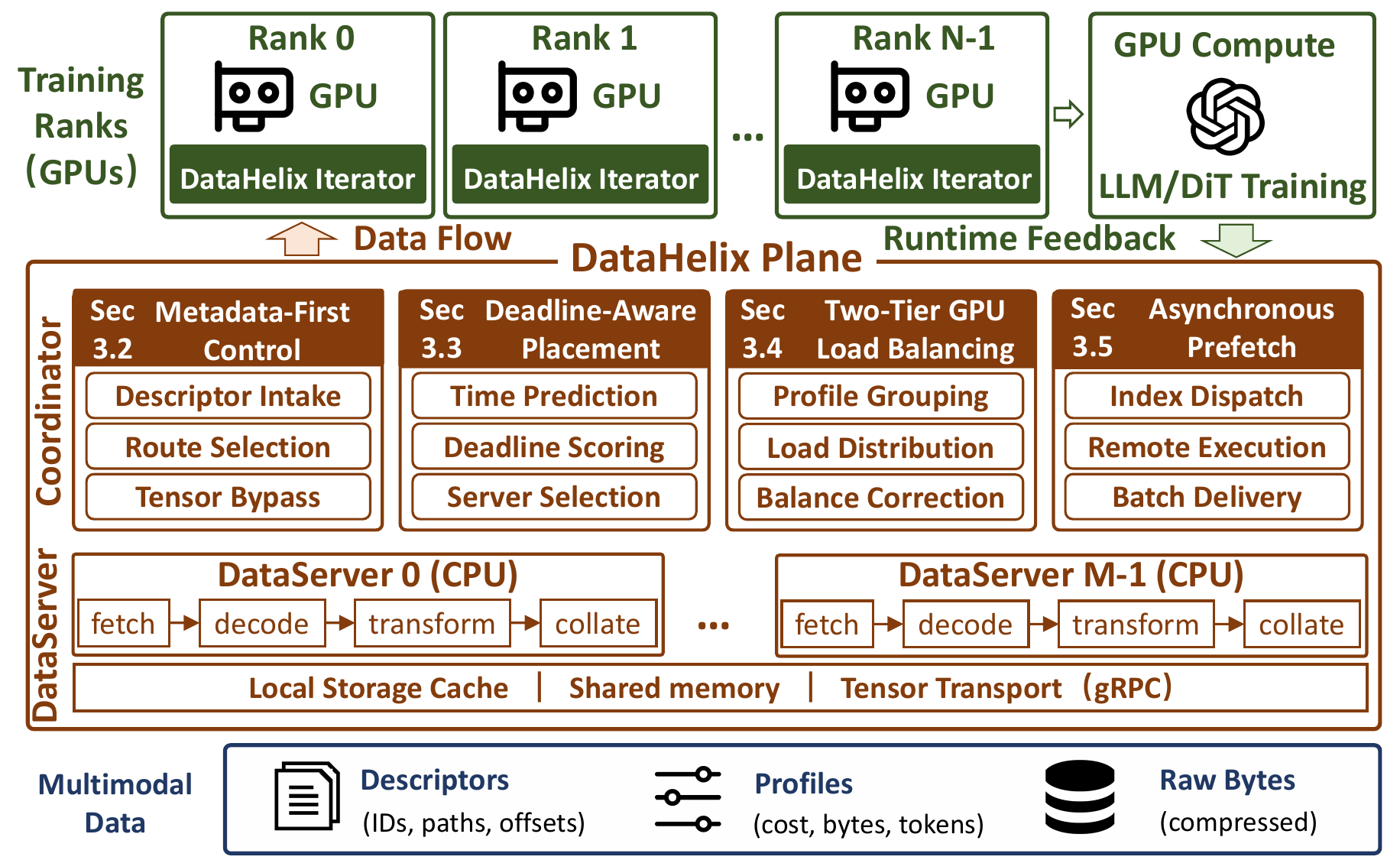}
  \caption{\systemname design: metadata-only scheduling with DataServer preprocessing and tensor delivery.}
  \vspace{1mm}
  \label{fig:block_diagram}
\end{figure}

\systemname turns the training input path from a collection of independent rank-local DataLoaders into a cluster-level data plane that uses GPU compute windows to set per-rank readiness targets. The core idea is to decide placement while a sample is still a lightweight descriptor and to keep heavyweight tensor payloads off the control path, so that CPU-side preprocessing can be scheduled to minimize per-rank deadline misses rather than global makespan, remote CPU sidecars can be exploited when they reduce deadline risk, and the training code still sees the ordinary DataLoader iterator interface.
Figure~\ref{fig:block_diagram} summarizes this philosophy. The Coordinator forms a metadata-only control loop that receives descriptor requests, uses cost and runtime feedback to choose an execution DataServer minimizing predicted deadline miss, and never touches decoded tensors. DataServers form the payload path: they execute dataset operations, exchange materialized batches via shared memory or HCCL/NCCL collectives when rebalancing is beneficial, and deliver batches toward the consuming rank's readiness target.


\subsection{Coordinator--DataServer Architecture}
\label{sec:design_arch}

\begin{figure}[t]
  \centering
  \includegraphics[width=\linewidth]{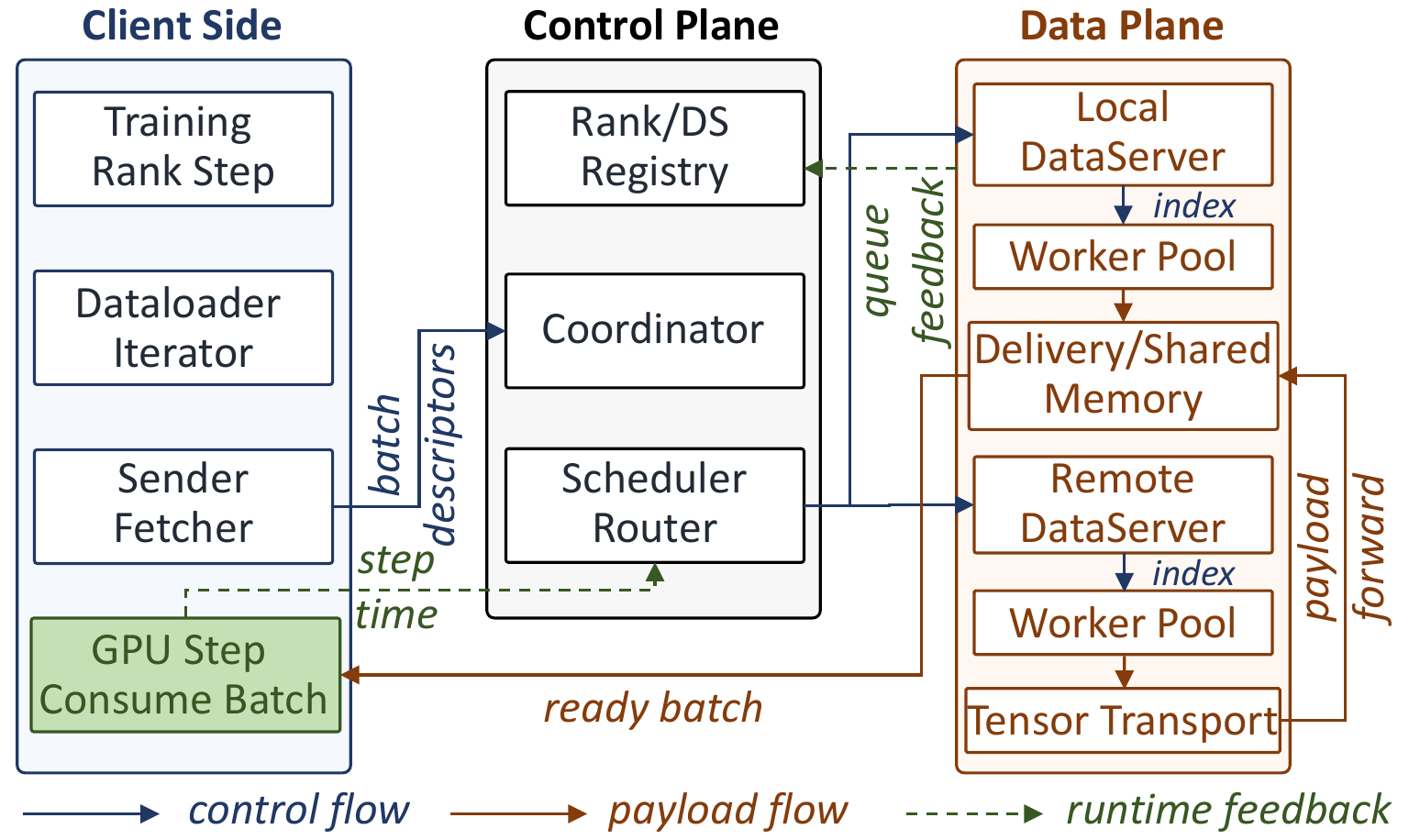}
  \caption{Runtime architecture: descriptor routing, direct payload delivery, and scheduling feedback.}
  \vspace{-1.5mm}
  \label{fig:minddata_arch}
\end{figure}

Like tf.data service~\cite{audibert2022tfdata_service} and MegaScale-Data~\cite{zhao2026megascale}, \systemname separates preprocessing from training. Figure~\ref{fig:minddata_arch} specializes this architecture for descriptor placement against per-rank readiness targets. The Coordinator observes DataServer load and runtime feedback but handles only metadata; DataServers execute preprocessing and deliver tensors directly. This separates iterator compatibility, placement control, and payload delivery without centralizing tensor traffic.

\textbf{Client Side.} Each rank keeps the familiar DataLoader iterator interface: a background sender issues descriptor-level batch requests ahead of the training thread, while a fetcher receives completed batches, preserves iterator order, and stages ready results for GPU consumption. The programming model is unchanged---the loop still calls \texttt{next(iterator)}---yet dispatch and result collection are no longer serialized.

\textbf{Control Plane.} The Coordinator maintains registered clients, local and remote sidecar DataServers, rank-to-DataServer mappings, queue indices, and heartbeat state. For each descriptor request it picks the execution DataServer via the policy of Section~\ref{sec:design_scheduler}; when that server is remote it attaches the target local DataServer to the descriptor so the materialized result reaches the right delivery endpoint. This gives the scheduler a global view of DataServer load and remote CPU availability without putting the Coordinator on the tensor-transfer path.

\textbf{Data Plane.} Each DataServer owns a worker pool running dataset fetch, video decode, image transforms, tokenization, collation, and exception propagation, with workers returning materialized results through shared-memory references. A local DataServer delivers directly to its co-located training rank, whereas a remote sidecar forwards the payload to the target local DataServer over the configured transport; when post-materialization rebalancing triggers, local DataServers use HCCL/NCCL collectives to redistribute at accelerator-interconnect bandwidth rather than CPU network speed. Heavyweight tensors thus bypass the Coordinator and move only between DataServers and training ranks.

This split has three advantages: placement decisions use cluster-level visibility rather than one rank-local queue; remote CPU capacity can be exploited when it reduces deadline risk while the common local path stays lightweight and free of a centralized payload bottleneck; and local DataServers leverage training-specific collective communication for high-speed rebalancing, cutting redistribution latency by 99\% versus gRPC-based alternatives.

\subsection{Metadata-First Control}
\label{sec:design_transport}

This layer optimizes decision timing and transfer volume. For video and multimodal samples, the cheapest moment to make a placement decision is before decode and collation, when the request is still a compact descriptor; once a tensor is materialized, load balancing may require moving tens or hundreds of megabytes across nodes. Metadata-first control therefore means more than sending metadata: it moves routing decisions before materialization and leaves tensor movement to the DataServer data plane only when needed.

\textbf{Descriptor intake.} The client sends only sampler indices and batching descriptors to the Coordinator. The DataLoader object is assigned lazily to a DataServer the first time that server executes the client's dataset, but the per-step request itself remains compact---important for multimodal workloads, where a video descriptor is a few bytes of indices and metadata while the decoded tensor can reach tens or hundreds of megabytes.

\textbf{Route selection.} For each descriptor request the Coordinator selects a DataServer before tensor materialization---balancing across the client's associated DataServers by default, scoring all candidates with the cost model of Section~\ref{sec:design_scheduler} when deadline-aware scheduling is on. If the selected server is not the requester's local DataServer, the Coordinator records the local target in the descriptor so the result reaches the right delivery point.

\textbf{Tensor bypass.} Batch payloads bypass the Coordinator. Locally, the DataServer delivers shared-memory metadata through a queue and the training process opens the referenced segment. Remotely, the execution DataServer extracts tensors from the Python batch structure, sends a compact structure description and tensor metadata, 
and transfers large buffers through the configured communication backend, with small tensors inlined and full gRPC serialization retained as fallback. The control path stays lightweight while decoded payloads use shared memory or peer transfer.

\subsection{Deadline-Aware Placement}
\label{sec:design_scheduler}

Latency-constrained scheduling is established in serving systems such as Clipper and Nexus~\cite{crankshaw2017clipper,shen2019nexus}. Here the target comes from the consuming training rank's GPU window: each descriptor is routed before materialization by comparing predicted batch-ready times against that target.


\textbf{Time prediction.}
For each candidate DataServer $i$, \systemname predicts when the requested batch would become ready if assigned there. The prediction is candidate-specific: a remote sidecar may have more idle CPU but adds forwarding cost, while a local DataServer avoids network transfer but can sit behind a deeper worker queue. Predicted finish time decomposes into CPU preprocessing, storage or input I/O, queueing delay, and materialized-payload transfer cost:

\begin{equation}
\label{eq:predicted_finish}
\begin{aligned}
T_i &= C_i^{cpu} + C_i^{io} + C_i^{queue} + C_i^{net},\\
M_i &= \max(0, T_i - D).
\end{aligned}
\end{equation}

These terms come from offline sample profiles and online heartbeat state: descriptor metadata such as frame count, resolution, input bytes, and locality hints give a pre-materialization estimate of $C_i^{cpu}$, $C_i^{io}$, and output size, while heartbeats refine $C_i^{queue}$ and $C_i^{net}$ from current queue depth, recent task latency, and peer-link bandwidth.

\textbf{Deadline scoring.}
In synchronous training, however, lowest predicted finish time is not the only objective: a candidate helps only if it can make the next batch ready inside the current GPU compute window. Given a data deadline $D$, the same scoring rule computes the predicted miss $M_i$ in Equation~\ref{eq:predicted_finish}.

Work stealing~\cite{blumofe1999scheduling} and LPT~\cite{graham1969bounds} provide established mechanisms for balancing work and reducing completion time. We specialize placement to synchronized consumption: the score combines predicted lateness, locality, and backpressure. When all candidates meet the readiness target, locality wins; when all miss, earliest-finish wins. This makes the cost of decoded-payload delivery part of the decision, rather than selecting a worker solely by queue length.

\textbf{Server selection.}
The final selection balances locality against deadline risk: if the local DataServer should meet the deadline, \systemname keeps the request local unless a remote candidate offers enough predicted gain, whereas a local path likely to miss makes a remote sidecar attractive despite the transfer cost. Online heartbeat feedback closes the loop—as queues grow or peer links congest, predicted finish times rise and future descriptors shift away from overloaded paths.

\subsection{Two-Tier GPU Load Balancing}
\label{sec:design_gpu_lb}

DynaPipe~\cite{jiang2024dynapipe} balances variable-length microbatch execution, while WLB-LLM and Zeppelin~\cite{wang2025wlbllm,chen2026zeppelin} coordinate computation and communication across parallel training. We apply cost-aware balancing to the input path in two stages (Algorithm~\ref{alg:two-tier-balance}). Tier 1 assigns lightweight descriptors using token/profile estimates before decoding. Tier 2 uses actual sample properties to correct residual skew only when projected savings justify moving decoded tensors. The DDLB runtime handles feature collection, assignment planning, and HCCL/NCCL redistribution; feature extraction and assignment policies remain customizable.

\begin{algorithm}[t]
\caption{Two-Tier GPU-Side Load Balancing.}
\label{alg:two-tier-balance}
\begin{algorithmic}[1]
\Require Sample set $\mathcal{S}$, rank set $\mathcal{R}$,
         cost profiles $\mathcal{P}$,
         DDLB feature function $f_{\mathrm{lb}}$,
         DDLB assignment policy $g_{\mathrm{lb}}$
\Ensure  Per-rank assignment $A: \mathcal{S} \to \mathcal{R}$ and
         redistribution plan $\Pi$

\Statex \textbf{// Tier 1: descriptor-level profile-aware assignment}

\State $\mathcal{G} \leftarrow \textsc{BucketByProfile}(\mathcal{S},\,\mathcal{P})$
       \Comment{group by $(\text{frame\_count} \times \text{resolution})$}
\State $\tau_r \leftarrow 0 \quad \forall\, r \in \mathcal{R}$
       \Comment{per-rank token accumulator}
\For{each profile group $G \in \mathcal{G}$}
    \For{each sample $s \in G$}
        \State $r^* \leftarrow \arg\min_{r \in \mathcal{R}}\; \tau_r$
               \Comment{greedy: route to least-loaded rank}
        \State $A(s) \leftarrow r^*$
        \State $\tau_{r^*} \leftarrow \tau_{r^*} +
               \mathcal{P}[s].\texttt{token\_estimate}$
    \EndFor
\EndFor

\Statex \textbf{// Tier 2: policy-driven post-materialization correction}

\State $\mathcal{B} \leftarrow \textsc{MaterializeBuffer}(A)$
       \Comment{small round of decoded samples}
\State $\mathcal{F} \leftarrow \textsc{AllGather}(f_{\mathrm{lb}}(\mathcal{B}))$
       \Comment{user-defined lightweight features}
\State $\widehat{A} \leftarrow g_{\mathrm{lb}}(\mathcal{F})$
       \Comment{custom policy may rebalance or keep current assignment}
\State $\Pi \leftarrow \textsc{PlanRedistribution}(A,\,\widehat{A})$
       \Comment{derive sends, receives, and keeps}
\State $A \leftarrow \widehat{A}$

\State \Return $(A,\,\Pi)$

\end{algorithmic}
\end{algorithm}

\textbf{Profile grouping.}
Sample count is a weak proxy for GPU work in video and multimodal training, so \systemname uses descriptor-level cost and profile information to build more representative batches before any tensor is materialized. For mixed video workloads the planner groups samples by profile—frame count and resolution—and preserves the intended mixture in the measured window, preventing a token-sorted prefix from containing mostly large clips.

\textbf{Load distribution.}
Within each global step the planner assigns profile-grouped samples to ranks while tracking per-rank token totals and per-profile counts---a zero-migration coarse balance that changes which descriptors each rank requests rather than moving decoded tensors afterwards. The same interface supports microbatch fanout, splitting one logical batch into smaller descriptor requests merged before the iterator yields.

\textbf{Balance correction.}
Descriptor-level planning cannot remove all skew, because real decode cost, augmentation behavior, and tensor sizes differ from static estimates. \systemname therefore adds a residual correction tier in the DataServer runtime---not merely a callback, but the DDLB execution path that buffers a small round of materialized samples, extracts lightweight features, gathers them across participating DataServers to turn an assignment decision into sends, receives, and keeps, and performs the exchange via the high-speed collectives of Section~\ref{sec:design_hccl_transfer}. The policy layer stays configurable---users supply the feature function and assignment policy while \systemname handles round coordination and redistribution---and in the current evaluation a cost gate keeps the current assignment unless the projected straggler reduction justifies post-materialization transfer, leaving profile-aware grouping and coarse rank balancing to the data-side batch planner.

\begin{lstlisting}[style=pythoncompact,caption={User-Defined DDLB function},label={load_balance_config}]
def sort_fn(features, args):
     if projected_gain(features) < args.min_gain:
         return keep_current_assignment(features)
     return rebalance_assignment(features)

DistributedDataLoader.launch(
     load_balance_config={
         "feature_fn": feature_fn,
         "sort_fn": sort_fn,
         "client_ranks": [[0, 1, 2, 3]],
         "buffer_size": 64,
         "args": policy_args,
     }
)
\end{lstlisting}

\subsection{High-Speed Redistribution via HCCL/NCCL}
\label{sec:design_hccl_transfer}

When post-materialization rebalancing triggers, the cost of cross-DataServer tensor transfer decides whether redistribution helps or hurts. Generic data planes use gRPC or TCP, which can take tens of seconds for gigabyte-scale decoded video tensors; \systemname instead leverages HCCL/NCCL—the same collective infrastructure used for gradient synchronization—to redistribute samples among local DataServers at accelerator-interconnect bandwidth.

\textbf{Dual-backend architecture.}
The main DataServer group uses gRPC with a CPU-based backend (gloo) for control-plane coordination: heartbeats, descriptor routing, remote sidecar integration. The load-balancing subgroup, containing only local DataServers, uses HCCL (Ascend) or NCCL (NVIDIA) for high-speed tensor redistribution, each local DataServer bound to a unique accelerator device so transfers proceed over NVLink, PCIe, or HCCS rather than the CPU network stack.

\textbf{Redistribution protocol.}
When Tier 2 rebalancing computes a new assignment $\widehat{A}$, the redistribution planner determines which samples each DataServer must send and receive. Because local DataServers share a common address space within the LB subgroup and each binds to a distinct device, \systemname uses collective \texttt{all\_to\_all} rather than point-to-point transfers: samples are packed into contiguous send buffers per destination rank and one collective completes the exchange in a single round.

\textbf{Impact.}
Redistributing 64 decoded video samples across 8 local DataServers takes roughly 74.5 seconds with gRPC-based one-by-one forwarding (the legacy fallback when HCCL initialization fails) but under 1 second with HCCL \texttt{all\_to\_all}—a 99\% reduction. The gain is training-specific, exploiting infrastructure already present for gradient synchronization rather than general-purpose CPU network optimization, and the dual-backend design keeps remote sidecars on gRPC while enabling high-speed local rebalancing when the projected benefit justifies movement.


\subsection{Asynchronous Prefetch}
\label{sec:design_prefetch}

\begin{figure}[t]
  \centering
  \includegraphics[width=\linewidth]{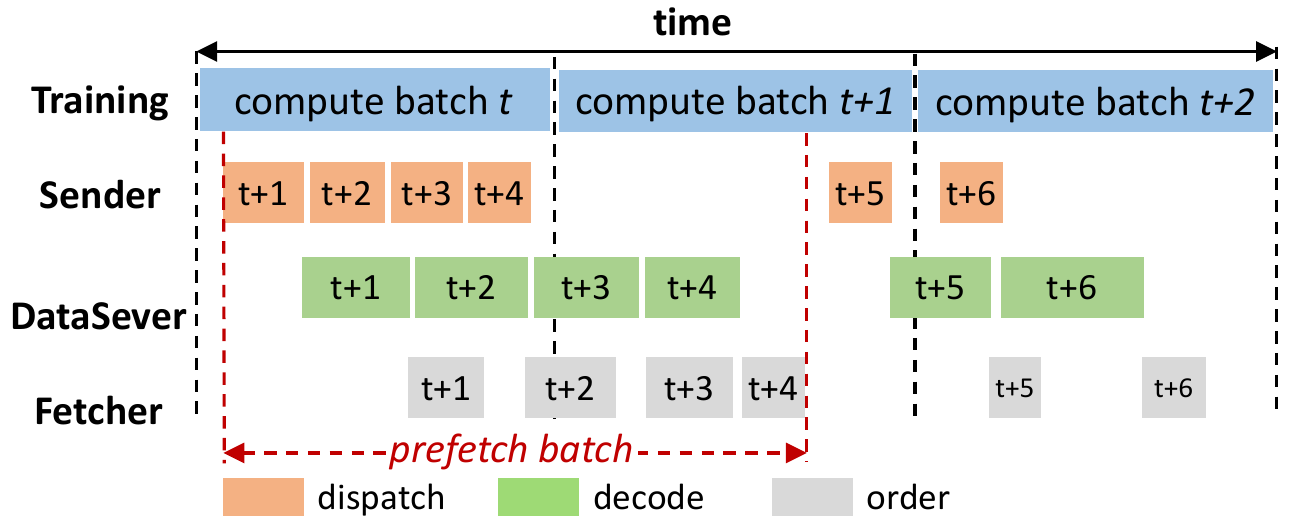}
  \caption{Asynchronous input preparation and ordered delivery overlap GPU execution (prefetch depth 4).}
  \label{fig:async_prefetch_timeline}
\end{figure}

This pipeline optimizes critical-path overlap while preserving iterator transparency. Even with good placement, the GPU can stall if descriptor dispatch, remote execution, tensor transfer, and batch assembly all start only after the training thread calls \texttt{next()}. \systemname therefore splits request submission from result consumption: the sender issues future descriptors ahead, DataServers prepare them while the current GPU step runs, and the fetcher restores ordered delivery to the standard DataLoader interface (Figure~\ref{fig:async_prefetch_timeline}).

\textbf{Index dispatch.}
A background sender thread reads the sampler, optionally splits a logical batch into microbatch descriptor requests, and sends them to the Coordinator subject to an in-flight limit. While the accelerator consumes batch $t$, descriptors for batch $t+1$ and beyond can already be queued at DataServers.

\textbf{Remote execution.}
Each DataServer enqueues assigned descriptors to its worker pool. Workers execute the dataset operation and return shared-memory-backed results to the dispatcher, which forwards them through the metadata-first tensor path when the request carries a remote target and delivers them locally otherwise.

\textbf{Batch delivery.}
A background fetcher thread consumes index responses and waits for the matching result from either the local DataServer queue or the gRPC data path. It preserves iterator order via task identifiers, optionally stages ready batches to the target device with non-blocking transfer, and places them in a client-side result queue. The training thread simply calls \texttt{next(iterator)} and receives an ordinary batch once the prefetcher finishes any required synchronization, so users keep the standard DataLoader programming model while \systemname pipelines descriptor dispatch, remote execution, tensor transfer, and batch delivery behind the current step.



\section{Implementation}\label{sec:optimizer}

We implement \systemname as an extension of the DataLoader execution path of PyTorch and MindSpore, adding global scheduling and remote preprocessing without requiring any changes to model training code.

The publicly shareable portion of the implementation is available in an anonymous repository~\cite{datahelix2026anonymous}.

\subsection{Programming Interface}
\label{sec:feature_api}


The \texttt{DistributedDataLoader} wrapper preserves the standard PyTorch
DataLoader iterator interface. Optional arguments configure asynchronous
prefetch, prefetch depth, microbatch fanout, and scheduler policy. Datasets can
expose descriptor-based costs through \texttt{estimate\_cost(index)}; otherwise,
\systemname uses conservative defaults.


\begin{figure}[t]
  \centering
  \includegraphics[width=\linewidth]{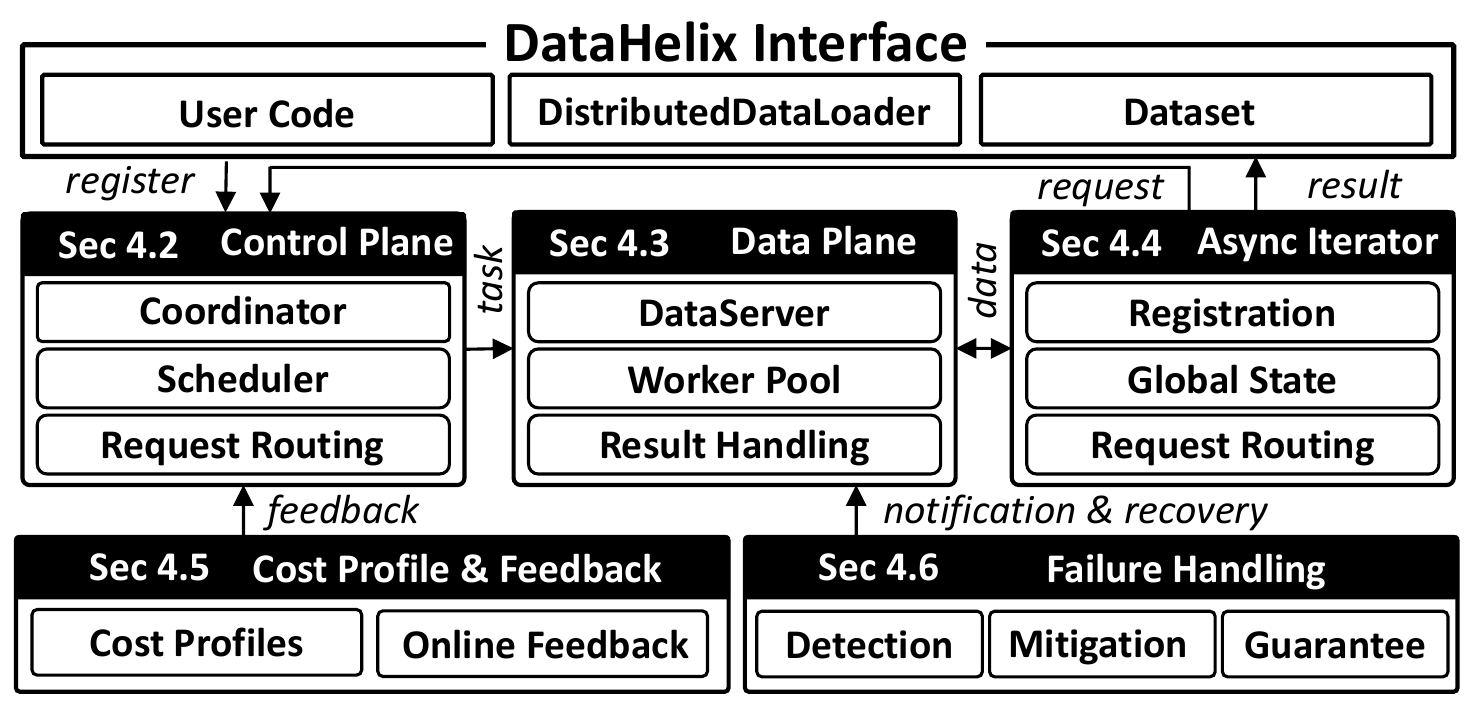}
  \caption{Overview of \systemname implementation.}
  \label{fig:implement}
\end{figure}

\subsection{Control Plane}
\label{sec:optimizer_control}

The Coordinator maintains global routing state in a gRPC service. Clients
register their rank ID and dataloader object; DataServers register their
address, rank, and heartbeat state. Requests use least-request routing by
default or the cost-aware scheduler in Sec.~\ref{sec:design_scheduler}.
For remote execution, each request carries the requesting rank's local
DataServer as its delivery target. The execution DataServer forwards the
materialized batch there, and the client response identifies this target,
preserving local queue/shared-memory delivery.

\subsection{Data Plane}
\label{sec:optimizer_dataplane}

Each DataServer dispatches a global task ID, serialized dataloader reference,
auto-collation flag, and index list to its fixed worker pool. Workers use the
ordinary dataset fetch path and serialize results into shared memory for
reconstruction or forwarding. Remote forwarding sends batch-structure metadata
and tensor descriptors first, with large tensors sent through the communication
backend when enabled; small tensors and fallback cases travel inline. The
target DataServer reconstructs batches into the client's local result queue.
Delivery follows client task IDs regardless of worker completion order;
worker errors propagate as iterator exceptions.


\subsection{Async Iterator Pipeline}
\label{sec:optimizer_prefetch_impl}

The pipeline in Sec.~\ref{sec:design_prefetch} uses an index-sender thread
bounded by the inflight limit and a fetcher thread that places ready batches
into the iterator queue. Training blocks only when this queue is empty.
Microbatch fanout stays within the client: it splits each logical batch into
ordinary index requests and waits for all parts before yielding the merged
batch, requiring no special Coordinator or DataServer protocol. Variable-shape
video tensors are padded before concatenation.

\subsection{Cost Profiles and Feedback}
\label{sec:optimizer_feedback}



Offline video profiles record decode profiles, estimated CPU time, encoded
input and decoded output sizes, and locality in a normalized JSONL table.
The Coordinator aggregates per-sample entries for each request. DataServer
heartbeats report worker capacity, task load and latency, and peer-link
bandwidth, latency, and inflight traffic to update queueing and network
estimates. Assignment telemetry records candidate scores, placement decisions,
reasons, and scheduling overhead for evaluating dispatch, cross-DataServer
transfer, and scheduler overhead.

\subsection{Failure Handling}
\label{sec:optimizer_failure}

Worker exceptions reach the training iterator so corrupt videos or missing
samples fail the responsible step instead of hanging the fetcher. Successful
benchmark runs write metrics, synchronize ranks, and write the rank-0 summary
before shutting down services, forcibly terminating them if graceful shutdown
times out; this cleanup does not affect the measured data path.

\section{Evaluation}\label{sec:evaluation}

We evaluate \systemname along three axes: a main study comparing a tuned native PyTorch DataLoader against \systemname data plane over the complete workload matrix; a controlled ablation isolating transport-stack separation, asynchronous prefetch with token-aware batching, and cost-aware DataServer scheduling; and a Ray actor pool baseline under the same CPU budget, testing whether a generic remote execution substrate can match \systemname with equal resources.

We begin with a latency breakdown to show the basic benefits and trade-offs of the data plane before presenting aggregate performance. We additionally examine whether the performance advantage persists as per-GPU batch size increases (Figure~\ref{fig:batch_size_scaling}).

\subsection{Experimental Setup}\label{sec:eval_setup}

\textbf{Platform.} Main experiments use one 8-card Ascend 910B server; multi-node experiments use $4 \times 2$ A100 GPUs. The native baseline launches eight active training ranks with rank-local DataLoader workers; \systemname also uses eight active training ranks but decouples data preparation into eight local DataServers plus a variable number of remote sidecar DataServers (denoted rd), sweeping rd8, rd16, and rd24. \systemname runs were validated to use the intended communication contract: local delivery uses the shared-memory/queue path, while remote tensor payloads use the communication path rather than a large gRPC blob fallback. 

\textbf{Equal CPU budget.} For the Ray baseline, we use the same e2e workload configuration and launch 96 Ray actors/CPU slots. \systemname rd16 uses 8 local and 16 remote DataServers with 4 CPU cores per DataServer, giving the same total budget of $(8+16)\times4=96$ CPU cores. Thus, the Ray comparison controls for provisioned CPU capacity while comparing the two data-pipeline designs on matched workloads. Equal budget refers to allocated CPU resources; CPU utilization and achieved throughput remain outcomes of each system.

\textbf{Workloads.} Table~\ref{tbl:pipelines_eval} summarizes the two workload classes. Video Training covers video model training on Kinetics and OpenVid, including Kinetics small, mid, large, and mixed decode profiles, where mixed alternates 16-frame 224p, 32-frame 336p, and 64-frame 448p samples. Multimodal SFT/PT covers video-text fine-tuning or pretraining with Qwen-VL-style models over Kinetics and OpenVid. The class label only groups modalities: results are reported per concrete workload rather than averaged within a class.


\begin{table}[t]
\centering
\caption{Evaluation workload classes and concrete workloads. Data-only experiments measure the data plane in isolation; e2e experiments additionally
include model forward, backward, and optimizer updates when a paired training model is available.}
\label{tbl:pipelines_eval}

\small
\setlength{\tabcolsep}{4pt}
\renewcommand{\arraystretch}{1.0}

\begin{tabular}{
@{}
L{0.18\linewidth}
L{0.14\linewidth}
L{0.55\linewidth}
@{}
}
\toprule

\multicolumn{1}{c}{\textbf{Class}}
&
\multicolumn{1}{c}{\textbf{Workload}}
&
\multicolumn{1}{c}{\textbf{Configuration}}
\\

\midrule

\multirow[c]{2}{*}{\makecell[l]{Video\\Training}}
&
Kinetics
&
S/M/L/Mixed video profiles; VideoMAE / Wan-1.3B
\\

&
OpenVid
&
Mixed video profiles; VideoMAE / Wan-1.3B
\\

\midrule

\multirow[c]{2}{*}{\makecell[l]{Multimodal\\SFT/PT}}
&
Kinetics
&
S/M/L/Mixed video-text data; Qwen-VL-3B SFT/PT
\\

&
OpenVid
&
Mixed video-text data; Qwen-VL-3B SFT/PT
\\

\bottomrule
\end{tabular}

\end{table}

\textbf{Metrics.} We report aggregate throughput (samples/s or videos/s), step latency, input-wait latency, DDL miss rate, scheduler overhead, dispatch breakdown, and cross-DataServer transfer volume. DDL miss rate is the fraction of measured steps whose input wait exceeds the configured data deadline. NPU utilization serves only as a secondary sanity check, since sampling can be sparse in short runs. For the 11-workload baseline summary, throughput uses a geometric mean across workloads; latency and DDL-miss metrics use arithmetic means. Each main-study cell uses 20 warmup and 100 measured steps, and the ablation follows the same convention with Kinetics mixed balanced across small, mid, and large samples.

\subsection{Latency Component Breakdown}\label{sec:eval_breakdown}

We first examine how the data plane changes latency in Qwen-VL training. We decompose rank-0 component summaries into data wait (request to local batch ready), compute (forward-backward, optimizer, and gradient synchronization), and synchronization wait, comparing the rank-local Native DataLoader, Ray, and \systemname rd16 on Kinetics-M and OpenVid-M.

\begin{figure}[t]
\centering
\includegraphics[width=\linewidth]{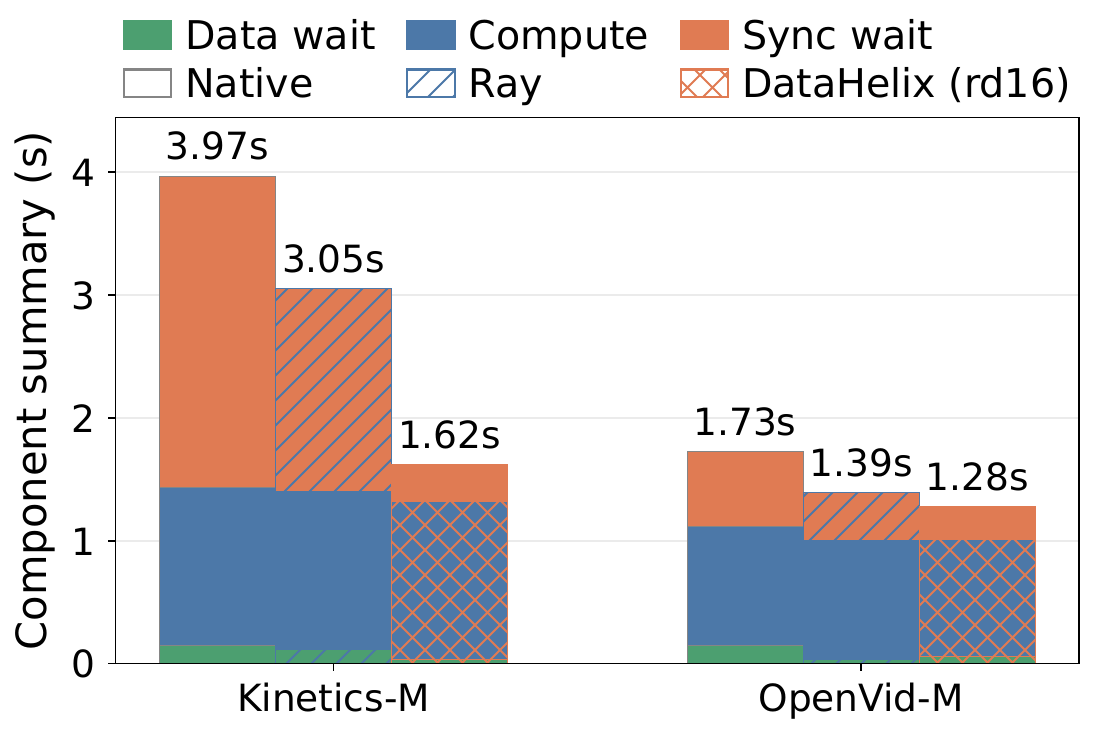}
\caption{Qwen-VL rank-0 latency summaries. Compute includes gradient sync; bar totals sum components, not measured end-to-end step latencies.}
\label{fig:latency_breakdown}
\end{figure}

Figure~\ref{fig:latency_breakdown} reports the three-component summaries for Native, Ray, and \systemname rd16. On Kinetics-M, \systemname rd16 reduces data wait from 0.147s to 0.037s and synchronization wait from 2.532s to 0.295s relative to Native, reducing the summed component time from 3.968s to 1.618s. Ray is intermediate on this workload (0.117s data wait, 1.652s synchronization wait, and 3.053s summed time), showing that matched CPU capacity alone does not reproduce the component profile of \systemname.

On OpenVid-M, the comparison exposes a trade-off rather than a uniform latency win. Ray has lower data wait (0.037s) than \systemname rd16 (0.055s), whereas \systemname has lower synchronization wait (0.270s versus 0.386s) and a lower summed component time (1.280s versus 1.391s). The basic benefit is therefore workload-dependent: \systemname reduces waiting on Kinetics-M and offers a different balance of latency components on OpenVid-M, without improving every component relative to Ray.

These rank-0 summaries diagnose component-level behavior; they are not cross-rank critical-path measurements. The bar-top labels sum the displayed component summaries and should not be interpreted as measured end-to-end step latency or used to infer throughput. We next assess end-to-end throughput and deadline compliance across the workload matrix.

\subsection{Main Performance Results}\label{sec:eval_main}

\begin{figure}[t]
  \centering
  \includegraphics[width=\linewidth]{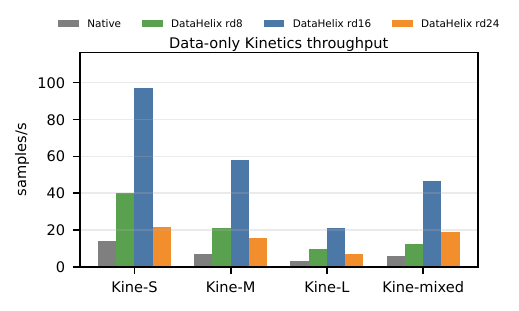}
  \caption{Data-only throughput on Kinetics profiles. }
  \label{fig:exp1_data_only}
\end{figure}

\begin{figure*}[t]
  \centering
  \includegraphics[width=0.95\textwidth]{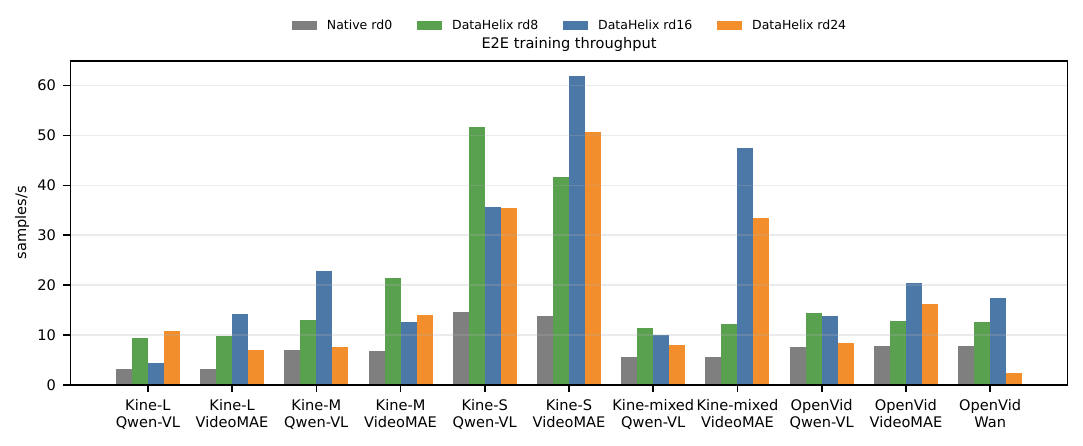}
  \vspace{-5mm}
  \caption{End-to-end throughput across workloads and remote DataServer counts.}
  \label{fig:exp1_e2e}
\end{figure*}

Figure~\ref{fig:exp1_data_only} isolates the data plane by replacing model training with a fixed synthetic window. On Kinetics, \systemname improves median data-preparation throughput by 2.85$\times$ at rd8, 7.45$\times$ at rd16, and 2.22$\times$ at rd24, peaking at 8.24$\times$ on Kinetics-mid: rank-local workers leave CPU parallelism unused, while a cluster-level DataServer pool absorbs decode, resize, and collation. The result is not monotonic in rd—rd16$\to$rd24 often lowers throughput, as extra remote workers add contention that offsets the CPU capacity.

Figure~\ref{fig:exp1_e2e} shows end-to-end throughput across the sweep. Restricting each setting to the 10 workloads with real Native and DataHelix results at rd8, rd16, and rd24, the median speedup is 2.53$\times$ at rd8, 2.53$\times$ at rd16, and 2.10$\times$ at rd24, peaking at 8.47$\times$ on Kinetics-mixed with VideoMAE at rd16. At rd16 Kinetics gains most (median 3.29$\times$) while OpenVid reaches 2.21$\times$ (maximum 2.61$\times$)—expected, since Kinetics decode and resolution profiles create more CPU-side skew for remote preprocessing to hide.

\textbf{Remote DataServer scaling.} The sweep shows rd8 already produces broad e2e gains and rd16 usually gives the best balance, while rd24 can regress: OpenVid-mixed with Wan reaches 2.21$\times$ at rd16 but falls below the native baseline at rd24. A data plane should therefore not simply maximize remote workers; it must account for deadline slack, queue depth, payload size, and network contention.

\begin{figure*}[t]
  \centering
  \includegraphics[width=0.95\textwidth]{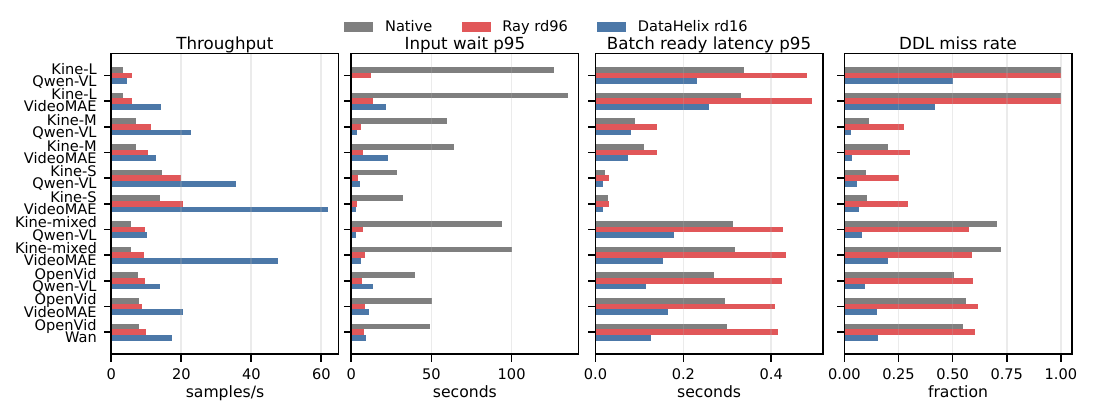}
  \vspace{-5mm}
  \caption{E2E baseline comparison across Native, Ray, and \systemname.}
  \label{fig:exp1_baseline_comparison}
\end{figure*}

\begin{table}[t]
\centering
\caption{Aggregate end-to-end comparison across 11 workloads. Throughput uses a geometric mean; other metrics use arithmetic means. Ratios are relative to Native. \textbf{Bold}/\underline{underline} values denote the best/2$^\mathrm{nd}$-best results, respectively.
}
\label{tbl:baseline_comparison_summary}

\small
\setlength{\tabcolsep}{4pt}
\renewcommand{\arraystretch}{1.0}

\begin{tabular}{@{}lccc@{}}
\toprule
\textbf{Metric} &
\textbf{Native} &
\textbf{Ray actor96} &
\textbf{\systemname rd16} \\
\midrule

\makecell[l]{Tput. GM$\uparrow$}
&
6.86 (1.00$\times$)
&
\underline{10.24 (1.49$\times$)}
&
\textbf{18.66 (2.72$\times$)}
\\

\makecell[l]{Wait p95$\downarrow$}
&
70.848s (1.00$\times$)
&
\textbf{7.902s (0.11$\times$)}
&
\underline{9.216s (0.13$\times$)}
\\

\makecell[l]{Ready p95$\downarrow$}
&
\underline{0.219s (1.00$\times$)}
&
0.311s (1.42$\times$)
&
\textbf{0.129s (0.59$\times$)}
\\

\makecell[l]{DDL miss$\downarrow$}
&
\underline{0.506 (1.00$\times$)}
&
0.554 (1.10$\times$)
&
\textbf{0.163 (0.32$\times$)}
\\

\bottomrule
\vspace{-5mm}
\end{tabular}
\end{table}

\textbf{Performance under equal CPU resources.}
Figure~\ref{fig:exp1_baseline_comparison} combines the two end-to-end baseline comparisons over 11 matched video and multimodal workloads across Native, Ray, and \systemname rd16 (the Kinetics-mixed/Qwen-VL row uses the user-confirmed historical supplement documented with the paper data); Table~\ref{tbl:baseline_comparison_summary} summarizes the set. Under the same 96-core budget, \systemname rd16 gains 1.82$\times$ geometric-mean throughput over Ray rd96, while relative to Native Ray reaches 1.49$\times$ and \systemname 2.72$\times$. The advantage over Ray persists after matching CPU capacity, so more CPU alone does not reproduce \systemname's performance.

The equal-budget advantage extends to batch readiness and deadline compliance: \systemname reduces mean batch-ready p95 from Ray's 0.311 s to 0.129 s and the mean DDL miss rate from 0.554 to 0.163. Input-wait tails trade off differently—Ray reaches a lower mean input-wait p95 of 7.902 s than \systemname's 9.216 s, though both beat Native's 70.848 s. The comparison thus establishes higher throughput and better deadline compliance at equal CPU budget without implying that every latency metric improves; the ablation below attributes the combined effect to individual components.

\begin{figure}[t]
  \centering
  \includegraphics[width=0.95\columnwidth]{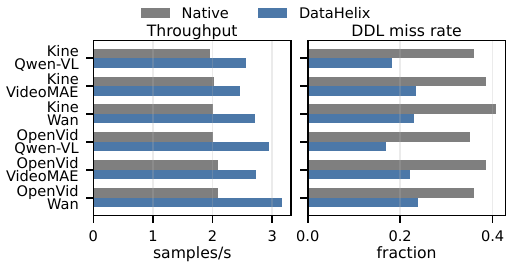}
  \caption{Multi-node throughput and DDL misses: Native versus \systemname on mixed-profile workloads.}
  \label{fig:multinode_native_minddata}
\end{figure}

\textbf{Multi-node Comparison.} Figure~\ref{fig:multinode_native_minddata} reports a follow-up multi-node comparison on Kinetics-mixed and OpenVid-mixed workloads. Across the 6 matched workloads, \systemname improves throughput over Native by 1.36$\times$ geometric mean while reducing the average DDL miss rate from 37.5\% to 21.2\%, indicating that the control/data-plane split remains useful when remote DataServers sit on separate machines rather than being emulated as sidecars.

\subsection{Batch-Size Scaling}\label{sec:eval_batch_scaling}

\begin{figure}[t]
  \centering
  \includegraphics[width=\linewidth]{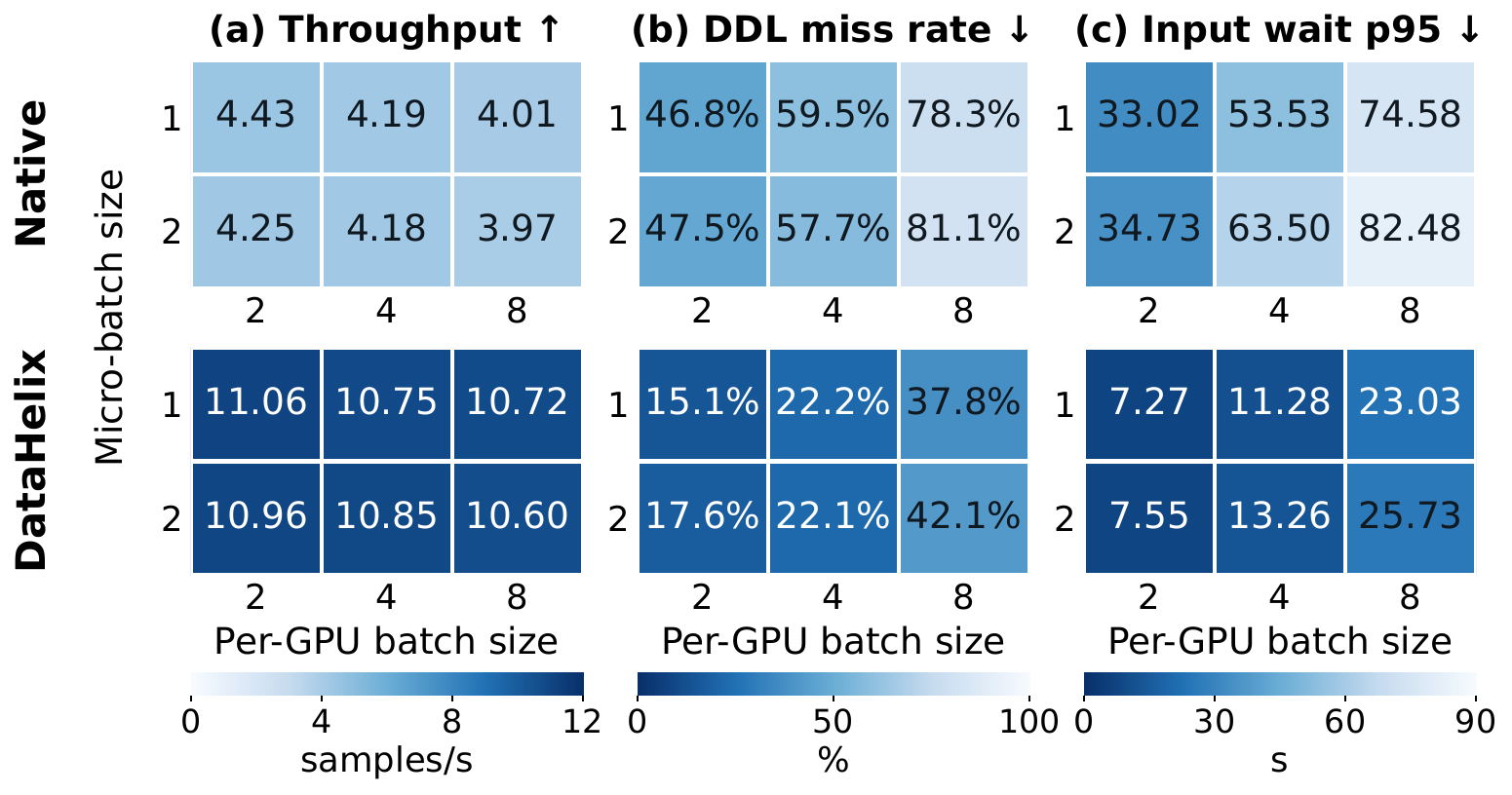}
  \caption{Batch-size scaling. Darker is better; each metric shares a color scale across methods.}
  \label{fig:batch_size_scaling}
  \Description{Heatmaps compare Native and DataHelix throughput, deadline miss rate, and input-wait p95 for per-GPU batch sizes 2, 4, and 8 and micro-batch sizes 1 and 2. DataHelix maintains nearly stable throughput as batch size grows and has lower deadline miss rates and input-wait p95 in every tested configuration.}
\end{figure}

We vary per-GPU batch size over $\{2,4,8\}$ and micro-batch size over $\{1,2\}$, using 40 warmup and 100 measured steps per configuration. The throughput heatmap in Figure~\ref{fig:batch_size_scaling}(a) shows that \systemname sustains its throughput advantage as batches grow. Increasing batch size from 2 to 8 retains 96.7--96.9\% of its throughput across the two micro-batch settings, compared with 90.6--93.4\% for Native. Its throughput advantage consequently grows from 2.50--2.58$\times$ at batch size 2 to approximately 2.67$\times$ at batch size 8.

Figure~\ref{fig:batch_size_scaling}(b,c) shows that larger batches increase DDL misses and input-wait tails for both systems, but \systemname remains substantially better in all six configurations. At batch size 8, its DDL miss rate is 37.8--42.1\%, versus Native's 78.3--81.1\%, and its input-wait p95 is 23.03--25.73\,s, versus 74.58--82.48\,s. Thus, over the tested batch-size range, \systemname maintains nearly stable throughput while limiting deadline misses and tail waiting relative to Native.

\subsection{Ablation Study}\label{sec:eval_ablation}

\begin{table}[!t]
\centering
\footnotesize
\caption{End-to-end ablation. \textbf{Bold}/\underline{underline}: best/second-best per workload, before rounding.}
\label{tab:ablation_scheduler_metrics}
\setlength{\tabcolsep}{2.5pt}
\begin{tabular*}{\linewidth}{
  @{\extracolsep{\fill}}l
  S[table-format=2.2]
  S[table-format=1.3]
  S[table-format=2.1]
  S[table-format=3.1]
  @{}
}
\toprule
\textbf{Feature} &
{\makecell[c]{\textbf{Tput.}$\uparrow$\\(samples/s)}} &
{\makecell[c]{\textbf{Input wait}$\downarrow$\\p95 (s)}} &
{\makecell[c]{\textbf{DDL miss}$\downarrow$\\(\%)}} &
{\makecell[c]{\textbf{Cross-DS}$\downarrow$\\(GB/100 steps)}} \\
\midrule
\multicolumn{5}{@{}l}{\textit{Video: Kinetics / Video-1B}} \\
Base DS          & 8.83 & {\underline{7.792}} & 18.9 & {\underline{199.6}} \\
+Transport       & 8.03 & 7.908               & 20.2               & 203.5 \\
+Prefetch/token  & {\underline{9.43}} & 7.877 & {\underline{18.9}} & {\bfseries 178.5} \\
+Cost/DDL        & {\bfseries 9.87} & {\bfseries 7.430} & {\bfseries 16.5} & 233.0 \\
\midrule
\multicolumn{5}{@{}l}{\textit{Video: OpenVid / Video-1B}} \\
Base DS          & 9.80 & 6.075 & 27.0 & 131.8 \\
+Transport       & 9.53 & 5.695 & 24.3 & {\underline{131.5}} \\
+Prefetch/token  & {\underline{11.07}} & {\underline{5.271}} & {\underline{19.6}} & {\bfseries 121.1} \\
+Cost/DDL        & {\bfseries 11.21} & {\bfseries 4.727} & {\bfseries 17.5} & 183.5 \\
\midrule
\multicolumn{5}{@{}l}{\textit{Multimodal: Kinetics / Qwen-VL-3B}} \\
Base DS          & 9.70 & 4.889 & 12.9 & {\underline{96.0}} \\
+Transport       & {\underline{10.82}} & 4.094 & 11.9 & {\bfseries 88.8} \\
+Prefetch/token  & 10.73 & {\bfseries 1.805} & {\underline{7.9}} & 106.4 \\
+Cost/DDL        & {\bfseries 11.13} & {\underline{2.267}} & {\bfseries 6.4} & 122.6 \\
\midrule
\multicolumn{5}{@{}l}{\textit{Multimodal: OpenVid / Qwen-VL-3B}} \\
Base DS          & 10.76 & 2.687 & 7.5 & 65.8 \\
+Transport       & {\underline{12.84}} & 2.769 & 7.7 & {\underline{62.8}} \\
+Prefetch/token  & {\bfseries 13.27} & {\underline{1.955}} & {\underline{6.0}} & {\bfseries 58.0} \\
+Cost/DDL        & 12.64 & {\bfseries 0.792} & {\bfseries 5.2} & 78.9 \\
\bottomrule
\end{tabular*}
\end{table}

The main study measures the full system; a feature-incremental ablation attributes the improvement to individual mechanisms. Its cumulative x-axis is: \emph{Base DS pool} (eight active training ranks with local and remote DataServers under least-request routing), \emph{+Metadata transport} (Queue/SHM local delivery, metadata-first remote forwarding, explicit task identity, CommunicationBackend tensor transfer), \emph{+Prefetch/token batch} (token/profile-aware batch planning, asynchronous prefetch, batch-internal microbatch fanout), and \emph{+Cost/DDL scheduler} (cost-profile and deadline-aware placement with runtime feedback). The reported ablation uses eight active ranks and evaluates W2 Video Training and W3 Multimodal SFT/PT on Kinetics-mixed and OpenVid-mixed.

\textbf{Prefetch/token batching improves video throughput.} Table~\ref{tab:ablation_scheduler_metrics} shows that prefetch/token batching produces the largest throughput increment for both video workloads, e.g., 8.03$\to$9.43 samples/s on W2 Kinetics. Multimodal gains differ: W3 OpenVid benefits most from transport, while W3 Kinetics gains little throughput from prefetch despite a sharply shorter input-wait tail. These differences reflect the balance between preprocessing, payload movement, and training compute.

\textbf{Cost/DDL scheduling targets deadline behavior and tail latency.} Compared with prefetch/token batching, Cost/DDL reduces DDL misses on all four workloads and input-wait p95 on three. For W3 OpenVid, DDL misses fall from 6.0\% to 5.2\% and input-wait p95 from 1.955 to 0.792 s. W3 Kinetics remains the exception: p95 rises from 1.805 to 2.267 s despite fewer misses. The scheduler therefore targets deadline compliance and tail latency, with workload-dependent throughput effects.

\textbf{The trade-off is higher cross-DataServer movement.} Cost/DDL increases Cross-DS transfer on all four workloads, e.g., 178.5$\to$233.0 GB/100 steps on W2 Kinetics. The weights of transfer cost, remote-placement gain, and DDL miss penalty control this trade-off between communication and deadline compliance.




\section{Related Work}\label{sec:related}

\subsection{ML Training Data Pipelines}

\textbf{Disaggregation and placement.} tf.data service~\cite{audibert2022tfdata_service} disaggregates preprocessing and coordinates reads to mitigate stragglers from unequal input sizes, while GoldMiner~\cite{zhao2023goldminer} elastically scales preprocessing workers. FastFlow~\cite{um2023fastflow} profiles local/remote execution and network limits to choose offloading decisions; Pecan~\cite{graur2024pecan} combines dynamic worker placement with transformation reordering to reduce cost. MegaScale-Data~\cite{zhao2026megascale} addresses multisource orchestration and parallelism-aware data delivery. These systems establish the value of global coordination, which our Coordinator--DataServer architecture also uses (Sec.~\ref{sec:design_arch}). Our placement decision operates on each descriptor request, combining predicted preparation and decoded-payload delivery time with the consuming rank's readiness target (Sec.~\ref{sec:design_scheduler}).

\textbf{Pipeline efficiency and readiness.} CoorDL~\cite{mohan2021coordl} reduces redundant I/O and preprocessing through coordination and caching; Cachew~\cite{graur2022cachew} combines caching with feedback-driven worker scaling. Plumber~\cite{kuchnik2021plumber} tunes parallelism, prefetching, and caching, while cedar~\cite{zhao2024cedar} optimizes operator execution across local and distributed resources. FFCV~\cite{leclerc2023ffcv}, DALI~\cite{nvidia2019dali}, and FusionFlow~\cite{kim2024fusionflow} accelerate data loading and transformations, with FusionFlow explicitly sharing preprocessing across CPUs and GPUs. These operator and resource choices affect the costs in Eq.~\ref{eq:predicted_finish}; our scheduler scores their sum against each rank's readiness target. MinatoLoader~\cite{nouaji2026minatoloader} tackles sample-time variability by constructing batches from readily prepared samples on one multi-GPU server. Our asynchronous iterator preserves task order (Sec.~\ref{sec:design_prefetch}), addressing readiness through advance dispatch and placement rather than completion-driven batch construction.

\subsection{Heterogeneity-aware Batch Scheduling}

Sequence packing~\cite{krell2021efficient} and length-based dataset decomposition~\cite{pouransari2024dataset} reduce padding or attention work. DynaPipe~\cite{jiang2024dynapipe} jointly plans variable-length microbatches and pipeline execution; WLB-LLM~\cite{wang2025wlbllm} and Zeppelin~\cite{chen2026zeppelin} balance model computation and communication through packing, sharding, or sequence partitioning. These approaches motivate using token/profile costs rather than sample counts, but their scheduling objects are model workloads. Our two-tier scheme (Sec.~\ref{sec:design_gpu_lb}) applies this principle before preprocessing and couples residual rank balancing to the cost of moving decoded inputs. The prefetch/token-batching ablation (Sec.~\ref{sec:eval_ablation}) evaluates the combined input-side mechanism.

\subsection{Scheduling Objectives and Granularity}

Work stealing~\cite{blumofe1999scheduling} balances task execution, and Ray~\cite{moritz2018ray} supplies distributed task and actor abstractions. \systemname specializes placement with descriptor costs and rank readiness; the CPU-matched Ray actor-pool comparison (Sec.~\ref{sec:eval_main}) tests the integrated design at equal provisioned capacity.
Latency-aware scheduling is also established: Clipper~\cite{crankshaw2017clipper} and Nexus~\cite{shen2019nexus} batch inference under latency objectives, while Orca~\cite{yu2022orca} schedules generative inference at iteration granularity. Pollux~\cite{qiao2021pollux} instead co-adapts training configurations and cluster allocations for goodput. Our deadline is an input-readiness target within a training job, with overruns causing stalls. The Cost/DDL ablation (Sec.~\ref{sec:eval_ablation}) makes its consequence explicit: fewer deadline misses can require more cross-DataServer traffic, and need not improve every latency metric.

\section{Conclusion}

This paper presents \systemname, a CPU--GPU co-scheduling data plane that moves
placement decisions before tensor materialization.  Its PyTorch-compatible implementation combines descriptor-level planning, metadata-first routing,
DataServer-based preprocessing and delivery, asynchronous prefetching, and
cost/DDL-aware placement, allowing heterogeneous input work to be scheduled
against the GPU step window rather than only average worker throughput.
Beyond the benchmarks above, our design has been merged into the data engine of a widely-used open-source AI framework, integrating directly with its existing dataset interfaces and model toolkits.




\balance
\bibliographystyle{ACM-Reference-Format}
\bibliography{paper}

\end{document}